\newcommand{\del}{\partial}

\newcommand{\bN}{\mathbb{N}}
\newcommand{\bZ}{\mathbb{Z}}
\newcommand{\boxtt}[1]{\mbox{\small\texttt{#1}}}

\renewcommand{\epsilon}{\varepsilon}

\newenvironment{fitem}{%
\vspace{-2\parskip}\begin{itemize} \itemsep = -0.2em%
}{%
\end{itemize}
}

\newcommand{\bmp}{\rule{0pt}{1pt}\\[-15pt]{\tiny.\dotfill.}\\*[-3pt]}
\newcommand{\bmpi}{\begingroup\footnotesize}
\newcommand{\empi}{\endgroup\rule{0pt}{1pt}\\*[-3pt]}
\newcommand{\empim}{\endgroup\hspace{5pt}\mbox{\small\(\maltese\)}\rule{0pt}{1pt}\\*[-3pt]}

\newcommand{\empix}{\endgroup\rule{0pt}{1pt}\\[-3pt]}
\newcommand{\emp}{\rule{0pt}{1pt}\\*[-17pt]{\tiny.\dotfill.}\\*[-10pt]\rule{0pt}{1pt}}

\newenvironment{ctikzpicture}{%
\begin{equation}\begin{tikzpicture}}
{\end{tikzpicture}\end{equation}}

\newtheorem{thr}{Theorem}

\newtheorem{dfn}[thr]{Definition}

\documentclass[a4paper,12pt]{article}
\usepackage{amsfonts}
\usepackage{amssymb}
\usepackage{amsmath}

\usepackage{verbatim}
\usepackage{graphicx}
\usepackage[usenames,dvipsnames]{color}
\usepackage{url}
\usepackage{tikz}
\usetikzlibrary{calc}

\definecolor{green}{rgb}{0,0.75,0}
\definecolor{violet}{rgb}{0.8225,0.2855,0.5225}
\definecolor{pink}{rgb}{1,0.5,0.5}
\definecolor{orange}{rgb}{0.8,0.6,0.3}

\title{Functional Programming is Free.}
\author{{\small\emph{Francis Sergeraert}}}
\date{\footnotesize January 2015}

\begin{document}

\voffset=-2.5cm \hoffset=-0.9cm \sloppy


\maketitle

{\abstract{A paper~\cite{ckmvw} has recently been published. This paper is
faulty: 1) The standard requirements about the definition of an algorithm are
not respected, 2)~The main point in the complexity study, namely the
\emph{functional programming} component, is absent. The Editorial Board of the
SIAM JC had been warned a confirmed publication would be openly commented, it
is the role of this text\footnote{Kent Pitman informed me that the esoteric
form \texttt{\#'(lambda ...)} is no longer necessary in ANSI Common Lisp; the
text is updated to the simpler form \texttt{(lambda ...)}.}.}}

\section{History of the subject.}

I started the subject of \emph{Effective Homology} in 1984, trying first to
implement the Edgar Brown's ``algorithm''~\cite{browne1} computing the homotopy
groups of the finite simply connected simplicial sets. This idea was quickly
understood as ridiculous; thirty years later, there remains difficult to hope
for an actual implementation of this algorithm, much too complex.

Other methods were soon imagined to overcome the essential difficulty,
consisting in using the strange power of the \emph{functional programming}.
Julio Rubio's contribution was essential to make the project succeed: he
understood in 1989 that the best theoretical tool to organize the work is the
so-called \emph{Basic Perturbation Lemma}~\cite{shih,brownr}, more and more
often called now the \emph{Homological Perturbation Theorem}, you may question
\texttt{\small Google(Galway perturbation)} to see numerous examples.

Using Julio's idea, a program called EAT (Effective Algebraic Topology) was
soon obtained in 1990, mainly devoted to the computation of the homology groups
of iterated loop spaces, already allowing us to produce homology groups so far
unreachable by the ``standard'' methods.

In the years 1995-8, Xavier Dousson's thesis was devoted to the calculation of
homotopy groups; the same tools, \emph{functional programming} combined with
the \emph{homological perturbation theorem}, were successfully applied to the
so-called Whitehead and Postnikov towers, the most ``elementary'' methods to
compute the homotopy groups. Leading to the Kenzo program~\cite{dsrbss}. In
particular, Ana Romero extended the scope of the program to the classifying
spaces of non-commutative groups, detecting an error~\cite{romrub} in a
published paper, for a homotopy group of a relatively contorted simplicial set.

  These programs, once the basics of algebraic topology are understood, are
elementary and certainly not the best possible: it is well known the Adams
spectral sequence and all its derivatives are central in the subject and
efficient methods \emph{must} use them. Nevertheless, the experience of the
Kenzo program, where everything is so elementary, made tempting a
non-negligible theoretical result: If a dimension \(n\) is fixed, our algorithm
computing the homotopy group \(\pi_n X\) of a simply connected simplicial set
\(X\) has a polynomial complexity \emph{with respect to} \(X\); of course not
with respect to \(n\), as it is well known since David Anick's
paper~\cite{anick}, unless \(P = NP\).

  This polynomiality result is so natural when you know the details of the
Kenzo program, and of course the authors of the corresponding 16000 lines of
Lisp code know these details, that I often stated this result without any other
justification. The authors of~\cite{ckmvw}, after a few years of interesting
common work with myself on problems of constructive algebraic topology,
suggested to write down a common paper on this polynomiality result. Why not.

  First Jiri Matousek identified a real difficulty that I had not seen: the
program rests on the effective homology of the Eilenberg-MacLane spaces, itself
based on the effective homology of \(K(\bZ,1)\), quite simple, but the standard
effective homology of this monstrous version of the circle \(S^1\) has
\emph{exponential} complexity! Jiri solved the problem for \(K(\bN,1)\), using
a lovely discrete vector field, and I gave the complement to obtain the same
result for \(K(\bZ,1)\), essentially an elementary process of normalization.
This is explained in~\cite{krmtsr}.

  No other obstacle against polynomiality was later found in the Kenzo
algorithm where everything is elementary. The paper~\cite{ckmvw} is a rereading
of the Kenzo source, or rather of the Kenzo documentation~\cite{siret}, with
some mathematical complements most often not included in the documentation. As
it often happens in journalistic reports, the very nature of the Kenzo program
is unfortunately in~\cite{ckmvw} misunderstood. The authors have an excuse: it
is hard to understand the core of such a program if you never worked with its
source code, or if you never did some similar concrete programming work.

Everything is elementary in Kenzo, except the functional programming component,
to my knowledge so far never so essentially and intensively used in a computer
program processing a \emph{specific} mathematical problem. All the
\emph{general} programs ``doing mathematics'', mainly computational algebra,
such as Maple, Mathematica, Axiom, Macsyma, Gap, Sage,~\dots also intensively
use functional programming, it is inevitable, but the corresponding complexity
problems are \emph{outside} the scope of functional programming. To my
knowledge, in our case, it is the first time an essential component of
functional programming is used in a program doing a work as specific as
computing a homotopy group.

  It so happened the authors of~\cite{ckmvw} in 2012 started preparing pages and pages
about our problem of complexity. I soon warned them: sorry, your draft totally
forgets the component \emph{functional programming}; without the minimal
indications about this matter, the paper is not correct. For example, a Coq
proof of this polynomiality result could not avoid studying the cost of this
functional programming. It was proposed I cosign the paper~\cite{ckmvw}, I
answered it is not possible to cosign a paper you would reject as referee.

Every mathematican knows how it is difficult in conversations with
non-mathematicians to explain what mathematical research is, and also why it is
useful~(\(!\)). Sometimes the same when a computer scientist tries to explain
to ``pure'' mathematicians the exact nature of a computer program; and in
particular where the difficulties are hidden. Modern programming, in particular
functional programming with the so-called \emph{functional programming
languages}, seems so natural for a pure mathematician, so close to his habits
that he cannot believe some difficulty in complexity is hidden there. Finally
it is not so hard to overcome this difficulty, but a few prerequisites cannot
be avoided. It is the role of this text to clarify this point.

\section{Functional Programming.}

It is common in Mathematics to study functions taking as arguments one or
several functions, the value of which being also a function. For example, an
elementary exercise of topology could be: Let \(\mathcal{F} :=
\mathcal{C}([0,1], [0,1])\) be the space of the continuous functions \(f: [0,1]
\rightarrow [0,1]\) with the usual metric topology; prove the function \(F:
\mathcal{F} \rightarrow \mathcal{F} : f \mapsto (f \circ f)\) is continuous.
The argument is \(f\), a function, the value is \(f \circ f\) a function, and
the problem concerns \(F\), a function: function \(\mapsto\) function.

We want to study in this text the analogous situation in computer science. It
is not difficult in most programming languages to write some code for functions
having as value a function; in programming, we say such a function
\emph{returns} a function, to emphasize the dynamic character of the process:
before been invoked, the ``value'' does not yet exist; when the dominating
function is \emph{invoked}, one says also it is \emph{called}, its work starts
and when the work is finished, a new object is born in the environment, a
\emph{functional object}, the object which is \emph{returned}. This functional
object may later in turn be invoked with appropriate arguments, returning
arbitrary objects, maybe again other functional objects, why not.

A programmer using such programming tools uses \emph{Functional
Programming}\footnote{This expression ``Functional Programming'' is also used
for a strict style of programming where \emph{every} (\(!\)) object is a
function, the simplest example being the \(\lambda\)-calculus, the wonderful
theoretical machine invented by Alonzo Church to disprove Hilbert's hope for a
general algorithm solving any mathematical problem. More reasonably, most
often, only the dynamic process defining how the program works then has a
functional nature.}. Functional programming is more or less easy, depending on
the programming language the programmer is using; some languages have been
specifically organized in particular to make easy this style of programming,
such a language then deserves the qualifier \emph{functional}. The main
functional languages are those of the Lisp family, those of the ML family and
Haskell.

  Algebraic Topology is often considered as a difficult subject, and our
experience allows us to point out the main reason causing this difficulty. With
only pen and paper, you cannot reasonably practice functional programming, it
is much too complex. A complexity which \emph{begins} to be illustrated by the
few diagrams of this text, used to explain the most elementary techniques which
are necessary when implementing functional programming. So that an algebraic
topologist wishing to work only with pen and paper must design quite
sophisticated methods to overcome this essential difficulty or more precisely
to use different strategies which avoid this obstacle: the exact and spectral
sequences are \emph{not} constructive; they become constructive only if you add
an essential dose of functional programming in the recipes. Most often, our
topologists must invent new high level exact sequences, spectral sequences,
spectra, operads, \ldots, made of higher homology or homotopy operators, also
of exotic spaces, with a \emph{constructive} scope always limited, a scope
which of course does not lower at all the interest of all these objects. While
our methods using functional programming have a unique constructive limitation,
the available material resources (computing time and memory space).

\section{Mathematics and Computer Science.}

It is not exceptional to observe mathematicians having a knowledge of computer
science which is a little superficial. This is true for any domain of
mathematics as well, but a mathematician often is not conscious of his exact
status with respect to computer science, the source of many problems.

Two extreme behaviours: some of them think a computer can only make essentially
trivial computations, which ``of course'' cannot do any relevant work for
``theoretical'' Mathematics. I remember a high level topologist who, sincerely
and honestly, declared he could not imagine a computer can determine whether a
natural number is prime or not. This happened frequently in the last decades of
the 20th century; many impressive results have been produced since then with
the help of computers and it becomes difficult to find extreme examples of this
sort.

The opposite extreme, now more frequent, is the following. Some mathematicians
think a computer is of course trivially able to implement any mathematical
idea, and some of them are surprised when their idea, a little naive, in fact
do not correspond to anything concretely feasible on a machine. Or is feasible
but using non-trivial implementations, the study of which then needing
significant and interesting work, in particular to justify \emph{theoretical}
mathematical results, for example in \emph{complexity}.

In a discussion with an excellent topologist, he explained a theoretical
algorithm for computing homotopy groups is known for a long time: it is enough
to use the \emph{combinatorial} definition of the homotopy groups by Kan and
Moore. He ``only'' forgot a finitely generated subgroup of a non-commutative
free group of finite type is free, but not necessarily of finite type and the
claimed ``algorithm'' is not valid.

Many topologists for example are persuaded the \emph{exact} and \emph{spectral
sequences} of Algebraic Topology can be trivially transformed into algorithms
computing some homology or homotopy groups. A referee recently claimed that
deducing an algorithm from the Adams spectral sequence is ``folklore'', sic. I
naively thought the mathematicians are the scientists the most careful for the
precise \emph{definition} of the objects they work with; it happens an
\emph{algorithm} is a non-trivial mathematical object with a deterministic
definition, due to the best representatives of our profession in the last
century: G\"odel, Church and Turing, and such a definition cannot be reduced to
some intuitive ``folklore''. As long as the referee reports will be
\emph{anonymous}, such accidents will be frequent: Juridically, an expert
cannot be anonymous, why this exception for the alleged ``experts'' judging our
articles?

This text is devoted to clarifying the particular case of \emph{functional
programming}. On the one hand, coding functions returning functions is
\emph{now} as easy in programming as in mathematics, mainly if you use a
so-called \emph{functional} programming language. On the other hand, the cost
in complexity is essentially \emph{null}, which is summarized by the title:
Functional Programming is \emph{free}; good news, but why? Because of the
subtle notion of \emph{closure}; of course the notion of closure in
programming, unknown by most mathematicians and even by a large proportion of
computer scientists.

Some complex programs intensively use functional programming, and studying
their complexity needs a lucid knowledge of this wonderful tool, the art of
producing \emph{closures}. Working in this way, you obtain efficient programs,
the structure of which is limpid; furthermore without any cost in complexity.
It is the subject of this article to explain this notion of closure, not so
complicated, but not trivial at all.

\section{Defining an algorithm.}\label{wi}

The paper~\cite{ckmvw} in principle proves some algorithm \(\pi_n: X \mapsto
\pi_n X\) is polynomial. Before proving this property, the algorithm \(\pi_n\)
must be available.

The proof of a \emph{theorem} is decomposed in simpler statements, often called
\emph{propositions}, maybe in turn decomposed in still simpler statements,
often called \emph{lemmas}, and these lemmas should have simple proofs directly
understandable. A similar scheme is to be applied for the definition of an
\emph{algorithm}.

The execution of a program finally is sequential, instruction by instruction,
so that the description of an algorithm consists in dividing the long sequence
of instructions into partial segments, divided in turn in shorter segments, and
so on, up to sufficiently elementary segments the implementation of which is a
routine task. Often, smaller segments consist in fact in a call of a functional
object done with respect to some prepared inputs, the caller waiting for the
corresponding output returned by the functional object. This is the standard
top-down view of algorithms, dual of the bottom-up understanding, the latter
easier to handle when writing source code.

The article~\cite{ckmvw} uses the Postnikov tower. The Whitehead tower is a
little simpler, produces also the homotopy groups, but not the \emph{Postnikov
classes}\footnote{Often erroneously called Postnikov \emph{invariants}.}. The
difficulty with respect to functional programming is the same for the Whitehead
tower and this framework is enough for our explanations.

The process computing \(\pi_n X\) thanks to the Whitehead tower consists in
iteratively constructing the stages of the tower, so that the first level
decomposition of the desired algorithm \(X \mapsto \pi_n X\) is necessarily:
\[
X = X_2 \stackrel{w_3}{\longmapsto} X_3 \cdots X_{n-1}
\stackrel{w_n}{\longmapsto} X_n \stackrel{H_n}{\longmapsto} \pi_n X
\]
The stage \(X_i\) of the tower is the inital space \(X\) where the first
homotopy groups \(\pi_j X\) have been ``killed'' for \(j < i\), its homotopy
starts at the index~\(i\). The Hurewicz theorem therefore allows the program to
compute \(H_i X_i = \pi_i X_i = \pi_i X\), and an appropriate fibration using
this group then produces the next stage \(X_{i+1}\). The last step \(H_n =
\pi_n\) is again an application of the Hurewicz theorem. The program starts
with \(X = X_2\) simply connected, without any homotopy in dimensions 0 and~1.

An algorithm is a functional object ``\emph{input} \(\mapsto\) \emph{output}''
and the computer scientists know for a long time the appropriate \emph{types}
of the input and the output are to be carefully defined. I invite a reader
sufficiently patient to reach this part of this text to analyse the
article~\cite{ckmvw} with the Adobe searcher for the word ``type'': the reader
will see this matter of the (computer) \emph{types} of the used objects is
never considered. The reader must probably understand the definition of these
types is a minor subject: the slave who some day maybe will implement the
described algorithm should be able to find out these details.

This matter of \emph{type} is yet become central in the very \emph{foundations}
of mathematics, see the Wikipedia page ``Homotopy Type Theory''. A point which
does not seem really understood by our authors. If they had taken a little care
of this question of \emph{types}, they would have realized some functional
programming is inevitable in such a programming.

In the first level decomposition of the Kenzo algorithm above, the main step is
the partial algorithm \(w_i: X_{i-1} \mapsto X_i\). The input and the output
are a simplicial set \emph{with effective homology}. We must therefore consider
the (computer) \emph{type} of the simplicial sets with effective homology.

It is not the right place to define all the details of this quite complex
computer type, a large collection of functional objects defining a simplicial
set in general not of finite type, combined with a set of other functional
objects sufficiently rich to contain the \emph{homotopy type} of this
simplicial set: this is nothing but the key point of \emph{Effective Homology}.
To illustrate our subject we just consider the simplest component of such an
object, the face operator \(\del_{X}\) of a simplicial set \(X\).

Such a face operator is a functional object, and you must first define\ldots\
the type of its input and output. The type of the simplices of \(X\) can be
chosen as a dependent type \(S := \bN \widetilde{\times} S_n\) where, for every
integer \(n \in \bN\), the type \(S_n\) is the type of the \(n\)-simplices of
\(X\). For example for the standard presentation of \(K(\bZ,1)\), a simplicial
set \emph{not} of finite type even if we limit the dimension, \(S_n\) is the
type of the lists of length \(n\) made of integers. The face operator \(\del\)
is then a functional object:
\[
\del:  [\bN \widetilde{\times} S_n] \widetilde{\times} \bN_\ast \rightarrow
[\bN \widetilde{\times} S_n] : ((n,\sigma),i) \mapsto \del_i(\sigma)
\]
where \(\sigma\) is an \(n\)-simplex of \(X\) and \(i\) an index \(0 \leq i
\leq n\). For example for \(X = K(\bZ,1)\), a possible 4-simplex is \(\sigma =
(-19, -14, 8, 4)\) and:
\[
\del_{K(\bZ,1)}((4, (-19, -14, 8, 4)), 2) = (3, (-19, -6, 4))
\]

Let \(X_{i-1}\) be a possible \((i-1)\)-th stage of a Whitehead tower,
necessarily implemented as a \emph{simplicial set with effective homology}.
This object contains a large collection of functional objects and in particular
the face operator \(\del_{X_{i-1}}\). The step \(w_i\) of the program must then
construct \(X_i := w_i(X_{i-1})\), the output \(X_i\) being again a simplicial
set with effective homology, this object containing in particular a similar
collection of functional objects, in particular the face operator
\(\del_{X_i}\); the construction of \(\del_{X_i}\) depends of course on the
object \(\del_{X_{i-1}}\), but also of all the other components of \(X_i\), in
particular on the other functional components, according to a process
computationnally not simple: a collection of various Eilenberg-MacLane spaces
must in particular be constructed also as simplicial sets with effective
homology, each Eilenberg-MacLane space being again the output of a functional
subprogram.

The paper~\cite{ckmvw} does not consider at all the cost of the generation of
all these functional objects. It is a pity:
\begin{fitem}
\item On the one hand, this \emph{functional programming} is the heart of the Kenzo
algorithm: if such a programming tool was not necessary, the topologists would
have used such an algorithm for a long time.
\item On the other hand, the complexity cost of the dynamic generation of the functional objects
is in this particular case essentially null. The rest of this text is devoted
to this point.
\end{fitem}%

\section{A toy example.} \label{compose} 

It could be interesting in a program, given two functional objects \(f_1, f_2:
\bN \rightarrow \bN\), to be able to dynamically construct the composition
\(f_2 \circ f_1: \bN \rightarrow \bN\), to be used later for some reason.

The Common Lisp style is convenient to illustrate such a situation. Let us
consider this short Lisp session.

 \bmp
 \bmpi\verb|> (setq f1 (lambda (n) (* n 3)))|\empim
 \bmpi\verb|#<Interpreted Function (unnamed) @ #x20ec3b8a>|\empix
 \bmpi\verb|> (setq f2 (lambda (n) (+ n 8)))|\empim
 \bmpi\verb|#<Interpreted Function (unnamed) @ #x20eca732>|\empix
 \bmpi\verb|> (list (funcall f1 5) (funcall f2 5))|\empim
 \bmpi\verb|(15 13)|\empi
 \emp

The first statement defines a function labeled ``\boxtt{f1}'' mathematically
defined as \(f_1: n \mapsto 3n\); read: assign to the symbol \boxtt{f1} the
function defined by the \(\lambda\)-expression \(\lambda n\,(n\,*\,3)\); the
same for \(f_2: n \mapsto n+8\). The last statement constructs a list, its
first element is \(f_1(5)\), the second one is \(f_2(5)\). Notice the Lisp
notation \boxtt{(funcall f1 5)} (read: call the function located by the symbol
\boxtt{f1} for the input \boxtt{5}) for the usual mathematical notation
\(f_1(5)\).

The reader may wonder why the returned functions are ``unnamed''. In this
organisation the symbol \boxtt{f1} ``sees'' the functional object, while the
functional object does not see at all the symbol \boxtt{f1}. In particular,
different symbols can see the same function. More concretely, the symbol
\boxtt{f1} has a pointer toward the functional object just created, and on the
contrary, the functional object has no pointer toward \boxtt{f1}. A mysterious
string such as \boxtt{\#x20dbc44a} is the hexa machine address of the
functional object.

If we want now to work with \(f_3 := f_2 \circ f_1\), we can use the Lisp
statement:

 \bmp
 \bmpi\verb|> (setq f3|\empi
 \bmpi\verb|    (lambda (n)|\empi
 \bmpi\verb|      (funcall f2 (funcall f1 n))))|\empim
 \bmpi\verb|#<Interpreted Function (unnamed) @ #x20dda8d2>|\empix
 \bmpi\verb|> (funcall f3 5)|\empim
 \bmpi\verb|23|\empi
 \emp

But this is essentially \emph{static}, the process constructing \(f_3\) should
be repeated if another composition is desired. If a program must construct a
large number of compositions, it would be painful to rewrite the same sort of
statement for each composition. It is better to write a general composition
function:

 \bmp
 \bmpi\verb|> (setq compose|\empi
 \bmpi\verb|    (lambda (af2 af1)|\empi
 \bmpi\verb|      (lambda (n)|\empi
 \bmpi\verb|        (funcall af2 (funcall af1 n)))))|\empim
 \bmpi\verb|#<Interpreted Function (unnamed) @ #x20e9bdf2>|\empix
 \bmpi\verb|> (setq f4 (funcall compose f2 f1))|\empim
 \bmpi\verb|#<Interpreted Closure (unnamed) @ #x20d9e092>|\empix
 \bmpi\verb|> (funcall f4 5)|\empim
 \bmpi\verb|23|\empi
 \bmpi\verb|> (setq f5 (funcall compose f1 f2))|\empim
 \bmpi\verb|#<Interpreted Closure (unnamed) @ #x20ea93aa>|\empix
 \bmpi\verb|> (funcall f5 5)|\empim
 \bmpi\verb|39|\empi
 \emp

You can read: the function \boxtt{compose} is the function which, given the
functions \boxtt{af2} and \boxtt{af1}, returns the function which, given the
integer \boxtt{n}, returns \(\boxtt{af2}(\boxtt{af1}(\boxtt{n}))\).

We have for example chosen the symbol \boxtt{af1} as vaguely meaning ``abstract
function \#1'' to improve the readability, but any arbitrary symbol could have
been chosen here, even the symbol \boxtt{f1}. The qualifier ``abstract'' means
that no concrete function is to be defined when the value of \boxtt{compose} is
\emph{constructed}; only later, when the functional object is \emph{used}, then
the abstract function will be replaced by some ``concrete'' function, as you
observe in the next statements.

But the most important point in this short Lisp session is the following: the
values of the constructed compositions \boxtt{f4} and \boxtt{f5} are
\emph{closures}, not functions. What does this mean? It is the role of this
text to clarify the exact nature of such an object, to illustrate the amazing
power of these objects, on the one hand in flexibility, on the other hand their
\emph{cost in complexity} is essentially \emph{null}.

\section{Functional objects.}

Except if you work with exotic machines, in particular machines allowing to use
multiprocessing, a program is finally a sequence of \emph{machine
instructions}, sequentially executed, except when some specific instruction, a
\emph{jump} instruction, says the next instruction to be executed is to be
taken at some machine address; the execution then again continues sequentially
from this address up to the next jump, and so on.

A program can also \emph{invoke} a functional object, this functional object is
then called a \emph{callee}, while the program at the origin of this call is
the \emph{caller}. The callee must then execute some task, and when this task
is finished, the \emph{control} is given back to the caller. The scheme below
illustrates such an organization; each rectangle contains some machine
instruction, and the arrows show the path followed by the \emph{control}. Such
a call consists in two jumps, one starting the callee, the other one leaving
it.

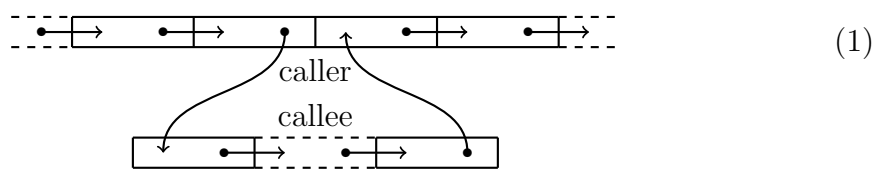
\begin{ctikzpicture} [thick, xscale = 0.4, yscale = 0.4, baseline = -1.5]
 \foreach \j in {0,1}
   {\draw (0,\j) -- (16,\j) ;
    \draw [dashed] (-2,\j) -- (0,\j) (16,\j) -- (18,\j) ;}
 \foreach \i in {0,4,...,16}
     {\draw (\i,0) -- (\i,1) ;}
 \foreach \j in {-4,-3}
   {\draw (2,\j) -- (6,\j) (10,\j) -- (14,\j) ; \draw [dashed] (6,\j) -- (10,\j) ;;}
 \foreach \i in {2,6,10,14}
     {\draw (\i,-4) -- (\i,-3) ;}
 \node [anchor = -90] at (8,-3) {callee} ;
 \node [anchor = 90] at (8,0) {caller} ;
 \foreach \i/\ii in {-1/1,3/5,11/13,15/17}
   {\draw [->] (\i,0.5) node {\scriptsize\(\bullet\)} -- (\ii,0.5) ;}
 \foreach \i/\ii in {5/7,9/11}
   {\draw [->] (\i,-3.5) node {\scriptsize\(\bullet\)} -- (\ii,-3.5) ;}
 \draw [->] (7,0.5) node {\scriptsize\(\bullet\)} to [out = -90, in = 90] (3,-3.5) ;
 \draw [->] (13,-3.5) node {\scriptsize\(\bullet\)} to [out = 90, in = -90] (9,0.5) ;
\end{ctikzpicture}%

In fact the caller, before invoking the callee, pushes onto the \emph{stack}
the \emph{return address}; in this way, when the callee has finished its task,
it can read this return address to give back the control to the caller, at the
next point where the caller must continue its work; the functional object can
so be called from anywhere in the program, avoiding to repeat its code in the
memory if this functional object is called several times, quite frequent.

A part of the computer memory is called the \emph{stack}; it is a continuous
area of the memory, where objects are \emph{piled}, the system being allowed to
add a new object onto it, or on the contrary to remove the last one having been
added; a special system register always keeps the address of the last object.
In particular, each time a functional object is launched for execution by a
caller, the caller adds (pushes) the return address onto the stack, so that
when the work of this functional object is finished, this address on the stack
is used by the callee to return to the caller at the right place, and this
return address is erased (popped); we will see this stack will be used also for
another specific usage of the closures, quite essential.

A \emph{closure} is also a functional object, but with a sophisticated
structure opening interesting new possibilities, in particular making easy
functional programming. A closure is \emph{abstractly} made of three
components:

\begin{fitem}
\item the \emph{body} of the closure, containing mainly:
  \begin{fitem}
  \item a pointer to its \emph{code};
  \item a collection of pointers defining its \emph{own} environment.
  \end{fitem}%
\item the \emph{code} of this closure;
\item the \emph{environment} of this closure.
\end{fitem}%

The \emph{apparent} redundancies in this description are voluntary. In a sense,
the wonderful power of our computers is due to the intensive and remarkable use
of the \emph{pointers}; a pointer is a machine object giving ``only'' the
machine address of another machine object; and it is not so rare the last
machine object is again a pointer, this is \emph{indirect addressing}, the key
point to implement the notion of closure (and many others).

When the functional object \emph{closure} will be invoked, its code will be
executed, but a part of this code may use the specific environment of
\emph{this} closure, the pointers available in the body itself allowing the
code to reach the relevant part of the current environment.

There remains to detail how this game of pointers is really organized, and why
this organization justifies the title of this text.

\section{The standard toy example.}\label{toy}

The standard mini-example which is used to explain the nature of the closures,
quite artificial but instructive, is the following. Let us imagine a programmer
wants to construct several arbitrary \emph{multipliers}, a multiplier \(\mu_m\)
being a functional object ready to multiply an arbitrary input \(n\) by some
fixed number \(m\). In standard programming, you could write:

 \bmp
 \bmpi\verb|> (setq mul3 (lambda (n) (* n 3)))|\empim
 \bmpi\verb|#<Interpreted Function (unnamed) @ #x20d50b7a>|\empix
 \bmpi\verb|> (funcall mul3 5)|\empim
 \bmpi\verb|15|\empi
 \emp

But our programmer intends to be able to \emph{dynamically}, that is, during
\emph{runtime}, \emph{generate} multipliers \(\mu_m\) with various and
arbitrary \(m\)'s, all these multipliers remaining simultaneously ``alive'',
and in particular such that they could be repeatidly called in arbitrary
orders. We assign a \emph{generator} of multipliers to the symbol
\boxtt{genmul}, and immediately use it:

 \bmp
 \bmpi\verb|> (setq genmul|\empi
 \bmpi\verb|    (lambda (m)|\empi
 \bmpi\verb|      (lambda (n) (* m n))))|\empim
 \bmpi\verb|#<Interpreted Function (unnamed) @ #x20d9c84a>|\empi
 \bmpi\verb|> (setq mul5 (funcall genmul 5))|\empim
 \bmpi\verb|#<Interpreted Closure (unnamed) @ #x20da2252>|\empix
 \bmpi\verb|> (setq mul7 (funcall genmul 7))|\empim
 \bmpi\verb|#<Interpreted Closure (unnamed) @ #x20daf022>|\empix
 \bmpi\verb|> (funcall mul5 9)|\empim
 \bmpi\verb|45|\empix
 \bmpi\verb|> (funcall mul7 9)|\empim
 \bmpi\verb|63|\empi
 \emp

The reader is invited to ``think'' like Lisp when these statements are
processed. This goes as follows.

\begin{fitem}
\item \boxtt{(setq genmul ...)}~: Lisp understands the user wants to use the symbol
\boxtt{genmul} and in particular assigns to it the object defined by the
continuation of the text.
\item \verb|(lambda (m) ...)|~: This object is an ``ordinary'' functional
object, the input \emph{will} be some object denoted by \boxtt{m} in the source
code and returning some output, some object defined by the subsequent text.
Important: at this time this functional object \emph{does not} work, it is then
just \emph{generated} and installed somewhere in the memory space; it will be
called only \emph{later}. We will also detail \emph{later}, \emph{when} this
functional object is called, how it works.
\item It happens the next statement \boxtt{(setq mul5 ...)} now \emph{calls}
\boxtt{genmul}, more precisely the functional object assigned to it, the input
being 5. At this time, the binding ``\boxtt{m} \(\mapsto\) 5'' defined by the
parameter \boxtt{m} and the input 5 is installed in the environment, more
precisely in the \emph{local} environment of the functional object just
starting its work.
\item The \emph{body} \verb|(lambda (n) ...)| of our functional object is then
executed, which body asks again for the \emph{generation} of a functional
object, not its execution.
\item Lisp then ``knows'' this generation happens in a \emph{local}
environment, a local environment having in this case only the binding
``\boxtt{m} \(\mapsto\) 5''. Then Lisp generates the next functional object
\verb|(lambda (n) ...)|, \emph{saving} also the current local environment, that
is, in this case the binding ``\boxtt{m} \(\mapsto\) 5'', this environment
being made visible \emph{only} from the just created functional object. This
\emph{pair} made of a functional object \emph{and} an environment valid only
for this object is a \emph{closure}, notion detailed later. This closure is the
\emph{output} of \boxtt{genmul} for this invocation, which closure is assigned
to \boxtt{mul5}.
\item Same story for \boxtt{(setq mul7 ...)}: \emph{another} functional object is
generated, assigned to \boxtt{mul7}, and this object, \emph{another} closure,
sees an environment where, this time the only binding is ``\boxtt{m}
\(\mapsto\) 7''. This closure is assigned to \boxtt{mul7}.
\item
The value of \boxtt{mul5} is \emph{called} with the argument 9, in other words,
the binding ``\boxtt{n} \(\mapsto\) 9'' is installed. The value of \boxtt{mul5}
is a closure which \emph{still sees} the binding ``\boxtt{m}~\(\mapsto\)~5'',
so that the \emph{code} of our closure, the Lisp statement \boxtt{(* m n)}
returns~45.
\item
The value of \boxtt{mul7} is \emph{called} with the same argument 9, in other
words, the binding ``\boxtt{n} \(\mapsto\) 9'' is again installed. This value
is a closure, not the same as the previous one, which \emph{sees} the
\emph{different} binding ``\boxtt{m} \(\mapsto\) 7'', so that the \emph{body}
of this closure, the Lisp statement \boxtt{(* m n)} this time returns 63.
\end{fitem}%

These details are a little lengthy but necessary to avoid deep
misunderstandings. Usually the following ``magic'' explanation is given: a
closure \emph{keeps} the local environment which \emph{was} present at
\emph{generation time}. A little more precisely: a closure \emph{encloses} the
local environment existing at generation time, explaining the terminology.
Partially true, but in general not sufficient, and the actual process is more
powerful.

\section{Detailing the structure of the closure \texttt{\bf mul5}.}

When \boxtt{genmul} is \emph{called} to \emph{generate} the closure which will
be assigned to \boxtt{mul5}, Lisp does the following work. The local
environment of \emph{this} invocation of \boxtt{genmul} has only the binding
``\boxtt{m} \(\mapsto\) 5''. Having to generate the functional object
\texttt{(lambda (n) (* m n)}, Lisp \emph{generates} in fact only a \emph{small}
structure, the body of the closure to be generated, containing only three
components:
\begin{fitem}
\item
A specific label indicating the object so entitled is a \emph{closure};
\item
The address of the code \texttt{(lambda (n) (* m n)};
\item
The address of the machine word containing the value 5 for the symbol
\boxtt{m}, value valid only for \emph{this} closure.
\end{fitem}%

You understand that, when \boxtt{genmul} will be later called to generate a
second closure assigned to \boxtt{mul7}, \emph{another} structure of the same
sort is again generated, ``disjoint'' of the first one, with the same label,
with, very important, the \emph{same} address of the (same!) code, but the
third component will be this time the address of the value~7 valid for  this
closure.

You must carefully distinguish in this story the functional object which,
\emph{when} it is called, generates a closure:
\begin{verbatim}
     (lambda (m)
       (lambda (n) (* m n)))
\end{verbatim}%
from the \emph{code}  \texttt{(lambda (n) (* m n)} of the generated closure.
When the interpreter or the compiler processes the functional object assigned
to \boxtt{genmul}, it ``sees'' this (ordinary) functional object will have to
generate a closure. The compiler then generates somewhere the (interpreted or
compiled) code of the closure, and this \emph{unique} copy of this code
\emph{will} be used by \emph{every} closure generated by \boxtt{genmul}.

However, this code, this unique code, according to the closure for which it is
working, has to use \emph{different} values for \boxtt{m}, values defined
beforehand, when the closure \emph{was} generated; how this is possible?

The solution is another use of the \emph{stack}, already mentionned to keep the
\emph{return} address from a functional object. When Lisp executes
\boxtt{(funcall mul5 9)}, Lisp works as follows:
\begin{fitem}
\item
\boxtt{(funcall ...)} means the \emph{value} of the first argument \boxtt{mul5}
of \boxtt{funcall} is a functional object.
\item
Examining the nature of this functional object, in this case a \emph{closure},
Lisp pushes on the {stack} \emph{three} addresses:
\begin{fitem}
\item
The return address, which will be used as for an ordinary functional object;
\item
The address of the closure.
\item
The adress of the second argument \boxtt{9}.
\end{fitem}%
\item This done, Lisp reads the address of the relevant code of \emph{this} closure
and gives the \emph{control} to this code.
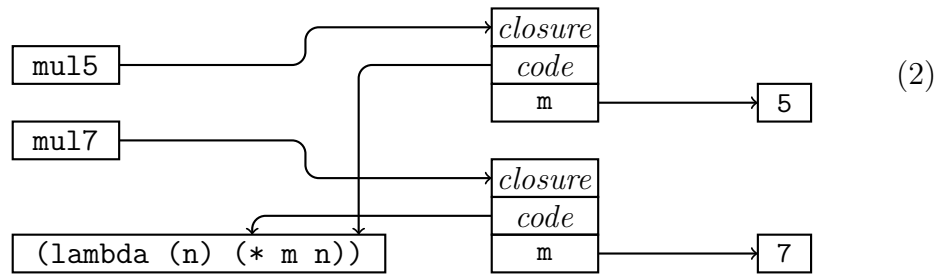
\begin{ctikzpicture} [thick, xscale = 0.7, yscale = 0.5, baseline = (base)]
 \coordinate (base) at (0,8) ;
 \draw (0,8) rectangle (2,9) ; \node at (1,8.5) {\texttt{mul5}} ;
 \draw (0,6) rectangle (2,7) ; \node at (1,6.5) {\texttt{mul7}} ;
 \draw (0,3) rectangle (7,4) ; \node at (3.5,3.5)
                                     {\texttt{(lambda (n) (* m n))}};
 \draw (9,9) rectangle (11,10) ; \node at (10,9.5) {\emph{closure}} ;
 \draw (9,8) rectangle (11,9) ; \node at (10,8.5) {\emph{code}} ;
 \draw (9,7) rectangle (11,8) ; \node at (10,7.5) {\boxtt{m}} ;
 \draw (9,5) rectangle (11,6) ; \node at (10,5.5) {\emph{closure}} ;
 \draw (9,4) rectangle (11,5) ; \node at (10,4.5) {\emph{code}} ;
 \draw (9,3) rectangle (11,4) ; \node at (10,3.5) {\boxtt{m}} ;
 \draw (14,3) rectangle (15,4) ; \node at (14.5,3.5) {\boxtt{7}} ;
 \draw (14,7) rectangle (15,8) ; \node at (14.5,7.5) {\boxtt{5}} ;
 \draw [->, rounded corners = 5pt] (2,8.5) -- (5.5,8.5) -- (5.5,9.5) -- (9,9.5) ;
 \draw [->, rounded corners = 5pt] (2,6.5) -- (5.5,6.5) -- (5.5,5.5) -- (9,5.5) ;
 \draw [->, rounded corners = 5pt] (9,8.5) -- (6.5,8.5) -- (6.5, 4) ;
 \draw [->, rounded corners = 5pt] (9,4.5) -- (4.5,4.5) -- (4.5, 4) ;
 \draw [->] (11,7.5) -- (14,7.5) ;
 \draw [->] (11,3.5) -- (14,3.5) ;
\end{ctikzpicture}
\item In other words this code is now executed.
\item
And when this code must use the \emph{value} of \boxtt{m} valid for \emph{this}
closure, the code has been prepared by the compiler to reach this value as
follows:
\begin{fitem}
\item
Read the address of the closure from the stack;
\item
Read the third component of the closure, which contains the \emph{address} of
the value of \boxtt{m} valid for \emph{this} closure.
\item
Finally read the value of \boxtt{m} valid for \emph{this} closure. Typical use
of \emph{indirect addressing}.
\end{fitem}%
\item
Via the stack, the code of our closure can also reach when necessary the second
argument \boxtt{9} of \boxtt{funcall}, that is, in this case, the unique
argument \boxtt{n} of the invoked closure.
\item When the execution of the code is finished, the code erases from the top
of the stack the address of the argument \boxtt{9} and the address of the
closure, no longer necessary, reads the return address on the stack, erases
also this return address, and is now able to \emph{return} to the caller at the
right place.
\item
Most often, a functional object must also return an \emph{output}, here
\boxtt{45}; the address of the output which has just been computed by the
closure is then put onto the stack, in this way the caller is able to reach
this output.
\end{fitem}%

Such a process:
\begin{fitem}
\item
Preparing some input for a callee;
\item
Giving control to this callee;
\item
The callee does its work;
\item
When the work is finished, the callee prepares the appropriate output;
\item
The callee gives back the control to the caller;
\item
All these steps using the stack to make the involved objects communicate
between each other;
\end{fitem}%
is so essential in any \emph{programming} language that the \emph{machine}
language has always specific instructions to conveniently and efficiently
handle the top of the stack as roughly described above. Examine for example the
entry ``Stack instructions'' in the Wikipedia page ``x86 assembly language''.
The particular case of the call of a closure is nothing but a variant of this
process where the address of the closure is also put onto the stack, to allow
the called code to reach the environment specific to this closure.

\section{A local environment is alive.} 

An acute reader maybe wondered why the closures above had their third component
containing the \emph{address} of the value of \boxtt{m} and not this value
itself, so needing an indirection.

But consider the next standard toy example:

 \bmp
 \bmpi\verb|> (setq make_package|\empi
 \bmpi\verb|    (lambda (m)|\empi
 \bmpi\verb|      (values|\empi
 \bmpi\verb|        (lambda (n) (* m n))|\empi
 \bmpi\verb|        (lambda () (setq m (1+ m))))))|\empim
 \bmpi\verb|#<Interpreted Function (unnamed) @ #x20db5a1a>|\empi
 \emp

The object assigned to the symbol \verb|make_package| is an ordinary functional
object using one argument \boxtt{m}. When this object is called, it returns
\emph{two} values, namely two closures. The first one is a multiplier \(\mu_m\)
as before. The second value is another closure which increases the value of
\boxtt{m} by 1. For example, we can call \verb|make_package| with the argument
``\boxtt{m} \(\mapsto\) \boxtt{5}''; the returned values are respectively
assigned to the symbols \verb|mul_1| and \verb|inc_1| as follows:

 \bmp
 \bmpi\verb|> (multiple-value-setq (mul_1 inc_1)|\empi
 \bmpi\verb|                       (funcall make_package 5))|\empim
 \bmpi\verb|#<Interpreted Closure (unnamed) @ #x20dba7e2>|\empi
 \emp

Only the first closure assigned to \verb|mul_1| is displayed. The same for two
other closures where now ``\boxtt{m} \(\mapsto\) \boxtt{18}'', and the symbols
\verb|mul_2| and \verb|inc_2| are used instead:

 \bmp
 \bmpi\verb|> (multiple-value-setq (mul_2 inc_2)|\empi
 \bmpi\verb|                       (funcall make_package 18))|\empim
 \bmpi\verb|#<Interpreted Closure (unnamed) @ #x20dbfb42>|\empi
 \emp

The reader probably guesses the continuation of the story:

 \bmp
 \bmpi\verb|> (funcall mul_1 -5)|\empim
 \bmpi\verb|-25|\empix
 \bmpi\verb|> (funcall inc_1)|\empim
 \bmpi\verb|6|\empix
 \bmpi\verb|> (funcall mul_1 -5)|\empim
 \bmpi\verb|-30|\empix
 \bmpi\verb|> (funcall mul_2 -5)|\empim
 \bmpi\verb|-90|\empix
 \bmpi\verb|> (funcall inc_2)|\empim
 \bmpi\verb|19|\empix
 \bmpi\verb|> (funcall mul_2 -5)|\empim
 \bmpi\verb|-95|\empix
 \bmpi\verb|> (funcall inc_1)|\empim
 \bmpi\verb|7|\empix
 \bmpi\verb|> (funcall mul_1 -5)|\empim
 \bmpi\verb|-35|\empi
 \emp

The new point is that both closures assigned to \verb|mul_1| and \verb|inc_1|
\emph{share} the \emph{same} ``local'' variable \boxtt{m}, for the environment
containing the initial binding was the \emph{same} for both closures. So that
once \verb|inc_1| has increased the value of \boxtt{m} from~5 to~6, the call of
\verb|mul_1| of course then uses the \emph{updated} value of \boxtt{m}. The
same for the closures assigned to the symbols \verb|mul_2| and \verb|inc_2|
with respect to their \emph{own} local environment, \emph{another} binding
``\boxtt{m} \(\mapsto\) \boxtt{18}'', updated to 19.

These complications do not prevent for example \verb|mul_1| and \verb|mul_2| to
share the same code. The appropriate scheme is the following one.

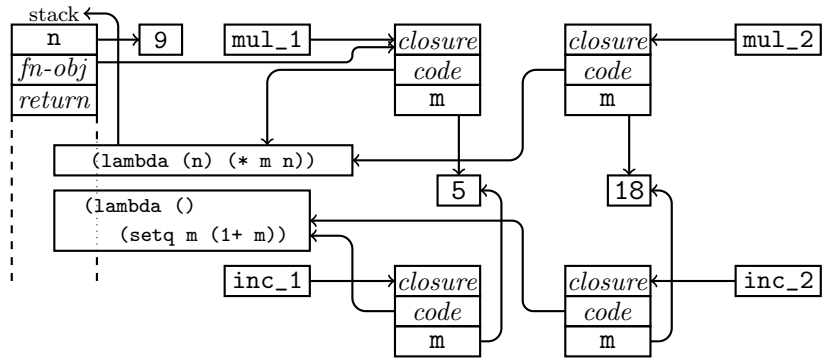
\begin{center}
\begin{tikzpicture} [thick, xscale = 0.56, yscale = 0.4, font = \footnotesize]
 \draw (5,10) rectangle (7,11) ; \node at (6,10.5) {\verb|mul_1|} ;
 \draw [->] (7,10.5) -- (9,10.5) ;
 \draw (5,2) rectangle (7,3) ; \node at (6,2.5) {\verb|inc_1|} ;
 \draw [->] (7,2.5) -- (9,2.5) ;
 \draw (17,10) rectangle (19,11) ; \node at (18,10.5) {\verb|mul_2|} ;
 \draw [->] (17,10.5) -- (15,10.5) ;
 \draw (17,2) rectangle (19,3) ; \node at (18,2.5) {\verb|inc_2|} ;
 \draw [->] (17,2.5) -- (15,2.5) ;

 \draw (9,10) rectangle (11,11) ; \node at (10,10.5) {\emph{closure}} ;
 \draw (9,9) rectangle (11,10) ; \node at (10,9.5) {\emph{code}} ;
 \draw [->,rounded corners = 5pt] (9,9.5) -- (6,9.5) -- (6,7) ;
 \draw (9,8) rectangle (11,9) ; \node at (10,8.5) {\boxtt{m}} ;
 \draw [->] (10.5,8) -- (10.5,6) ;
 \draw (9,2) rectangle (11,3) ; \node at (10,2.5) {\emph{closure}} ;
 \draw (9,1) rectangle (11,2) ; \node at (10,1.5) {\emph{code}} ;
 \draw [->, rounded corners = 5pt] (9,1.5) -- (8,1.5) -- (8,4) -- (7,4) ;
 \draw (9,0) rectangle (11,1) ; \node at (10,0.5) {\boxtt{m}} ;
 \draw [->, rounded corners = 5pt] (11,0.5) -- (11.5,0.5) -- (11.5,5.5) --
 (11,5.5);

 \draw (10,5) rectangle (11,6) ; \node at (10.5,5.5) {\boxtt{5}} ;
 \draw (14,5) rectangle (15,6) ; \node at (14.5,5.5) {\boxtt{18}} ;

 \draw (13,10) rectangle (15,11) ; \node at (14,10.5) {\emph{closure}} ;
 \draw (13,9) rectangle (15,10) ; \node at (14,9.5) {\emph{code}} ;
 \draw [->, rounded corners = 5pt] (13,9.5) -- (12,9.5) -- (12,6.5) -- (8,6.5) ;
 \draw (13,8) rectangle (15,9) ; \node at (14,8.5) {\boxtt{m}} ;
 \draw [->] (14.5,8) -- (14.5,6) ;
 \draw (13,2) rectangle (15,3) ; \node at (14,2.5) {\emph{closure}} ;
 \draw (13,1) rectangle (15,2) ; \node at (14,1.5) {\emph{code}} ;
 \draw [->, rounded corners] (13,1.5) -- (12,1.5) -- (12,4.5) -- (7,4.5) ;
 \draw (13,0) rectangle (15,1) ; \node at (14,0.5) {\boxtt{m}} ;
 \draw [->, rounded corners = 5pt] (15,0.5) -- (15.5,0.5) -- (15.5,5.5) --
 (15,5.5);
 \draw (1,6) rectangle (8,7) ; \node at (4.5,6.5)
                                     {\scriptsize\texttt{(lambda (n) (* m n))}} ;
 \draw (1,3.5) rectangle (7,5.5) ; \node at (3,5) {\scriptsize\texttt{(lambda ()}} ;
 \node at (4.5,4) {\scriptsize\texttt{(setq m (1+ m))}} ;
 \draw (0,10) rectangle (2,11) ; \node at (1,10.5) {\texttt{n}} ;
 \draw (3,10) rectangle (4,11) ; \node at (3.5,10.5) {\texttt{9}} ;
 \draw [->] (2,10.5) -- (3,10.5) ;
 \draw (0,9) rectangle (2,10) ; \node at (1,9.5) {\emph{fn-obj}} ;
 \draw [->, rounded corners = 2.5pt] (2,9.75) -- (8,9.75) -- (8,10.25) -- (9,10.25) ;
 \draw (0,8) rectangle (2,9) ; \node at (1,8.5) {\emph{return}} ;
 \draw [dashed] (0,2.5) -- (0,8) (2,7.2) -- (2,8) (2,2.5) -- (2,3.5) ;
 \draw [dotted, thin] (2,3.5) -- (2,7.2) ;
 \node [anchor = -90, inner sep = 1pt] (st) at (1,11) {\scriptsize stack} ;
 \draw [->, rounded corners = 5pt] (2.5,7) |- (st.0) ;
\end{tikzpicture}
\end{center}%

  The central components are the four closures, each one located by a symbol;
in this way, the programmer can call them easily. These closures share some
codes and share also some environments, but in this case not in parallel: if
two closures share a code, they do not share an environment, and conversely.
This, once the source code is interpreted or compiled, is \emph{static}.

In the diagram above, the state of the memory is also sketched when, a
\emph{dynamic} event, the closure located by \verb|mul_1| is called with the
argument \boxtt{9}. Then the control is given to the code of ``\boxtt{(lambda
(n) (* m n))}'', reached via the closure, which code can reach the binding
``\boxtt{m} \(\mapsto\) \boxtt{5}'' of the \emph{relevant} environment via the
stack and its \emph{fn-obj} component; the code can reach the binding
``\boxtt{n} \(\mapsto\) \boxtt{9}'' also via the stack.

  We hope this diagram could make clear the role of the closures: they are
very small boxes containing the address of the corresponding code and also the
addresses of the variables of the corresponding environment. \emph{When} a
closure is called, the control is given to the corresponding code, which
reaches the appropriate arguments via the stack and the closure, in particular
the variables of the closure itself.

  This is nothing but a \emph{toy example}. In more general situations, it is
quickly impossible to draw a planar diagram as above! Taking account of the
descriptive power of the addressing mechanism -- \(n\) bits can denote \(2^n\)
different addresses! --, this is not at all a problem for the interpreter or
the compiler. This is why, as it is detailed in the next section, functional
programming is essentially \emph{free}.

\section{When is a closure generated?}

We hope the reader is now sufficiently advanced in this matter of closures to
understand another point. Our reader could have been puzzled by the quoted
adjective ``ordinary'' often attributed to a functional object. How the
interpreter or the compiler can decide whether such an object must be
``ordinary'' or not, that is, in the negative case, a closure?

Some \emph{global} environment is always defined, that is, a collection of
bindings everywhere visible in the program, except when such a binding is
``hidden'' by a local one; on the contrary a \emph{local} environment is
visible only from a small (local) part of the code, \emph{and also} from the
closures which \emph{were} generated when this local environment was active.

When a functional object is to be generated, Lisp examines the state of the
environment where this generation is to be done. If only the global environment
is visible, an  ordinary functional object is generated; on the contrary, if
the generation happens inside some local environment, then a closure is
generated.

The simplest way in Lisp to generate a local environment uses a \boxtt{let}
instruction.

 \bmp
 \bmpi\verb|> (setq a 25)|\empim
 \bmpi\verb|25|\empix
 \bmpi\verb|> (let ((a 255))|\empim
 \bmpi\verb|     (+ a a))|\empi
 \bmpi\verb|510|\empix
 \bmpi\verb|> (+ a a)|\empim
 \bmpi\verb|50|\empi
 \emp

In this short session above, two uses of the variable \boxtt{a} coexist. The
global environment defines a binding ``\boxtt{a} \(\mapsto\) \boxtt{25}'',
while, when the \boxtt{(let ...)} instruction is run, a local environment with
the binding ``\boxtt{a} \(\mapsto\) \boxtt{255}'' is defined. In this local
environment, adding \boxtt{a} and \boxtt{a} produces 510, but this done, this
local environment is definitively dead, so that in the next statement, the same
code \boxtt{(+ a a)} returns this time 50: the global environment was let
unchanged and is reused as such.

Combining a \boxtt{(let ...)} statement with generation of functional objects
shows when the interpreter or the compiler decides to generate an ordinary
functional object or a closure.

 \bmp
 \bmpi\verb|> (setq adda1 (lambda (b) (+ a b)))|\empim
 \bmpi\verb|#<Interpreted Function (unnamed) @ #x20f10aaa>|\empix
 \bmpi\verb|> (setq adda2|\empi
 \bmpi\verb|    (let ((a 33))|\empi
 \bmpi\verb|      (lambda (b) (+ a b))))|\empim
 \bmpi\verb|#<Interpreted Closure (unnamed) @ #x20d88cba>|\empix
 \bmpi\verb|> (funcall adda1 66)|\empim
 \bmpi\verb|91|\empix
 \bmpi\verb|> (funcall adda2 66)|\empim
 \bmpi\verb|99|\empi
 \emp

The functional object assigned to \boxtt{adda1} is generated when only the
\emph{global} environment is visible, so that this object has only the status
of (ordinary) function. On the contrary, the object to be assigned to
\boxtt{adda2} is generated when the \emph{local} environment with the local
binding ``\boxtt{a} \(\mapsto\) \boxtt{33}'' is active, so that this requires a
closure saving this environment for later uses.

This effect is more visible when the global and the local environment are both
used in a closure.

 \bmp
 \bmpi\verb|> (setq a 11)|\empim
 \bmpi\verb|11|\empix
 \bmpi\verb|> (setq b 22)|\empim
 \bmpi\verb|22|\empix
 \bmpi\verb|> (setq add_a_b|\empi
 \bmpi\verb|    (let ((b 222))|\empi
 \bmpi\verb|      (lambda (n) (+ a b n))))|\empim
 \bmpi\verb|#<Interpreted Closure (unnamed) @ #x20e60bca>|\empix
 \bmpi\verb|> (funcall add_a_b 33)|\empim
 \bmpi\verb|266|\empix
 \bmpi\verb|> (setq a 110)|\empim
 \bmpi\verb|110|\empix
 \bmpi\verb|> (setq b 2)|\empim
 \bmpi\verb|2|\empix
 \bmpi\verb|> (funcall add_a_b 33)|\empim
 \bmpi\verb|365|\empi
 \emp

The closure uses the global binding ``\boxtt{a} \(\mapsto\) 11'' and the local
one ``\boxtt{b} \(\mapsto\) \boxtt{222}'' the last biding hiding the global one
``\boxtt{b} \(\mapsto\) \boxtt{22}''. The first call computes \(11 + 222 + 33 =
266\). Then the \emph{global} bindings of \boxtt{a} and \boxtt{b} are modified.
The second call of the closure computes \(110 + 222 + 33 = 365\); in other
words, the closure uses the unique (global) binding of \boxtt{a}, which has
been modified by \boxtt{(setq a 110)}, but the local binding of \boxtt{b} which
is not changed by \boxtt{(setq b 2)}, this last \boxtt{setq} changing only the
global binding of \boxtt{b}, which binding is hidden inside the local
environment of the closure by its own local binding, unchanged.

All the previous examples are artificial, just designed to make obvious how, in
a good programming language, the interpreter and the compiler process the
terrible problem of \emph{identifier scope} in a relatively sophisticated but
powerful way. This organization made of global environments, local environments
\emph{and closures} is the result of a long evolution of the art of
programming; the interested readers can consult the Wikipedia page ``Closure
(computer programming)'' to get some information about this evolution.

The author of this note does not forget his serious difficulties when he tried
in the 70's to understand the very notion of the \emph{funarg problem} in
Maclisp, see:

{\small\noindent\boxtt{http://www.maclisp.info/pitmanual/eval.html}}

\noindent when the birth of the notion of closure was just under way: the
programmer had to manage himself the problem of identifier scope through the
\emph{association lists}. The discussion described in this web page is
instructive to follow the hesitations of the best implementers of this time.

There remains to see why the use of these closures is \emph{free}, we mean in
complexity studies, and also why this organization is exactly the right one to
implement high level mathematics, we mean when notions coming from the
\emph{Category Theory}, typically the \emph{functors}, have to be implemented.

\section{Functional Programming is Free.}

We mean that, in a complexity study, the cost of the \emph{production} of a new
closure during the execution of a program is ``null'', more exactly is
\emph{constant}. In most complexity studies, the input \(\alpha\) of a program
\(\pi\) is assumed to have a \emph{size} \(\sigma(\alpha)\), the program is run
to produce some output \(\omega = \pi(\alpha)\), and we would like to bound the
computing time \(\tau(\pi, \alpha)\) in function of \(\sigma(\alpha)\). For
example, the program~\(\pi\) is said to be \emph{polynomial} if an exponent
\(e\) and a coefficient \(c\) can be determined, such that for any meaningful
input \(\alpha\), the estimate \(\tau(\pi, \alpha) \leq c(1 +
\sigma(\alpha))^e\) is satisfied.

Let us assume we would like to prove the program \(\pi\) is polynomial. If the
number of closures to be produced is bounded \emph{independently} of
\(\alpha\), the fact that the cost of the production of a closure is constant
implies the production time of these closures can be omitted: it is enough to
increase the coefficient \(c\) to take account of the production of closures.
Otherwise, when the number of closures depends on~\(\alpha\), such a hasty
claim could fail, but anyway using the fact the production cost of \emph{one}
copy of the closure is constant can be used conveniently.

Producing a closure of a certain type is only allocating a memory segment of
fixed length, a segment containing the label \emph{closure}, the address of the
corresponding code, always the same, and the addresses of a \emph{fixed} number
of local values.

Of course you must distinguish the computing time necessary to \emph{generate}
the closure from the computing time of the closure itself when it is
\emph{called} later by the program; this computing time can on the contrary
strongly depend on the inputs given by the caller to the closure. For example
the closure could be an Ackermann function depending on some parameters; the
production of the closure is then constant, but the computing time of such a
closure when it is \emph{called} of course is not constant at all, depending on
the input \emph{and also} possibly on the specific parameters of \emph{this}
Ackermann function.

The investigator in complexity must also be lucid about another factor. The
\emph{production} itself of a closure needs a constant time, but
\emph{preparing} the local environment valid for this closure can on the
contrary be costly, depending on the parameters of this closure. Let us
consider for example the case of a function depending on a natural number
\(n\), this function having to produce a closure which must use the \(n\)-th
prime number. It is then better to compute this prime number before generating
the closure, and to insert the result in the environment known by the closure.
In Common Lisp the structure of the function generating the closure could be:

 \bmp
 \bmpi\verb| > (setq closure_gen|\empi
 \bmpi\verb|     (lambda (n ...)|\empi
 \bmpi\verb|       (let ((nth_p (funcall nth_prime n)))|\empi
 \bmpi\verb|          ...|\empi
 \bmpi\verb|          (lambda (...) ... nth_p ...))))|\empim
 \emp

The (ordinary) function assigned to \verb|closure_gen| in particular depends on
the natural number \boxtt{n} and maybe on other arguments. The \boxtt{(let
...)} instruction inside the definition of the function enriches the current
environment with a binding ``\verb|nth_p| \(\mapsto\) \boxtt{13}'' if the
function is called with ``\boxtt{n} \(\mapsto\) 6''; in other words, a function
computing the \(n\)-th prime number is assumed assigned to the symbol
\verb|nth_prime|. This done, and maybe some other auxiliary work, the closure
is generated \emph{inside} the environment generated by the \boxtt{(let ...)};
the closure therefore will keep the address of the value of \verb|nth_p|, so
that the code of the closure will be able to directly use this value without
having to recalculate it. In such a situation, the cost of the calculation of
the \(n\)-th prime is to be added to the cost of the generation of the closure.

In other words, three steps when a closure is used:
\begin{fitem}
\item Preparing the appropriate local environment, the cost depends on the
specific necessary work;
\item Generating the closure itself, constant cost.
\item Using the closure when it is called, cost depending on the code of the closure,
on the arguments given by the caller, and also possibly of the nature of the
environment used by the closure.
\end{fitem}%

\section{Implementing functors thanks to appropriate closures.}

To explain why the notion of closure is the right tool to implement
\emph{functors}, we use the following elementary example. Let us assume a
programmer has to work with (the category of) monoids, and in particular, he
must implement the product \emph{bifunctor}.

He decides a monoid is a pair, the first element being the \emph{identity} of
the monoid, the second one being the \emph{operation} of the monoid. Such an
operation is a process, a function, which, given two elements of the monoid
computes their product. For example the additive monoid of the natural numbers
could be implemented as:

 \bmp
 \bmpi\verb|> (setq N+ (list 0 #'+))|\empim
 \bmpi\verb|(0 #<Function +>)|\empi
 \emp

The cryptic text \boxtt{\#'+} means in Lisp the functional object associated to
the standard addition operator \boxtt{`+'}. Test:

 \bmp
 \bmpi\verb|> (funcall #'+ 4 5)|\empim
 \bmpi\verb|9|\empi
 \emp

The \emph{value} of the expression \boxtt{\#'+} is \boxtt{\#<Function +>}, an
external simple representation of the functional object, too complicated  to be
exactly displayed, it is some machine code, readable only by experts.

Our programmer uses also the multiplicative monoid of the integers.

 \bmp
 \bmpi\verb|> (setq N* (list 1 #'*))|\empim
 \bmpi\verb|(1 #<Function *>)|\empix
 \bmpi\verb|> (funcall #'* 4 5)|\empim
 \bmpi\verb|20|\empi
 \emp

An \emph{identity} function is defined to legibly extract the identity of a
monoid.

 \bmp
 \bmpi\verb|> (setq identity|\empim
 \bmpi\verb|    (lambda (monoid)|\empi
 \bmpi\verb|      (first monoid)))|\empim
 \bmpi\verb|#<Interpreted Function (unnamed) @ #x20eafb12>|\empix
 \bmpi\verb|> (funcall identity N+)|\empim
 \bmpi\verb|0|\empi
 \emp

And an \emph{operation} function is also defined to conveniently compute
\emph{inside} some monoid; the second statement below constructs a list made of
the respective ``products'' of \boxtt{4} and \boxtt{5} in the monoids
\boxtt{N+} and \boxtt{N*}.

 \bmp
 \bmpi\verb|> (setq operation|\empi
 \bmpi\verb|    (lambda (monoid item1 item2)|\empi
 \bmpi\verb|      (funcall (second monoid) item1 item2)))|\empim
 \bmpi\verb|#<Interpreted Function (unnamed) @ #x20eb9cca>|\empix
 \bmpi\verb|> (list (funcall operation N+ 4 5)|\empi
 \bmpi\verb|        (funcall operation N* 4 5))|\empim
 \bmpi\verb|(9 20)|\empi
 \emp

Now we implement the product \emph{bifunctor}. If \(M_1\) and \(M_2\) are two
monoids, an element of \(M_1 \times M_2\) is represented as a pair. So that the
identity of \(M_1 \times M_2\) is the pair of the respective identities. The
function which, given the monoids \(M_1\) and \(M_2\), constructs the identity
of \(M_1 \times M_2\) therefore is:

 \bmp
 \bmpi\verb|> (setq monoid-product-identity|\empi
 \bmpi\verb|    (lambda (monoid1 monoid2)|\empi
 \bmpi\verb|      (list (funcall identity monoid1) (funcall identity monoid2))))|\empim
 \bmpi\verb|#<Interpreted Function (unnamed) @ #x20ec729a>|\empix
 \bmpi\verb|CL-USER(120): (funcall monoid-product-identity N+ N*)|\empim
 \bmpi\verb|(0 1)|\empi
 \emp

We did not yet have seen any closure. In the same way, we must be able to
construct, given two \emph{arbitrary} monoids \(M_1\) and \(M_2\), the
operation of the product monoid \(M_1 \times M_2\).

 \bmp
 \bmpi\verb|> (setq monoid-product-operation|\empi
 \bmpi\verb|    (lambda (monoid1 monoid2)|\empi
 \bmpi\verb|      (lambda (item1 item2)|\empi
 \bmpi\verb|        (list (funcall operation monoid1|\empi
 \bmpi\verb|                       (first item1) (first item2))|\empi
 \bmpi\verb|              (funcall operation monoid2|\empi
 \bmpi\verb|                       (second item1) (second item2))))))|\empim
 \bmpi\verb|#<Interpreted Function (unnamed) @ #x20ef3f92>|\empix
 \bmpi\verb|> (funcall monoid-product-operation N+ N*)|\empim
 \bmpi\verb|#<Interpreted Closure (unnamed) @ #x20ef9102>|\empi
 \emp

The operation to be constructed uses the operation of the first monoid and
makes it work on the first components of the ``items'', the same for the second
components with respect to the second monoid, and finally makes a pair, a list,
with both results.

Using this constructor with respect to our monoids \boxtt{N+} and \boxtt{N*}
this time produces a \emph{closure}. The point is that the internal
\boxtt{(lambda (item1 ...) ...)} uses \boxtt{monoid1} and \boxtt{monoid2}
constituting the \emph{local} environment where the functional object is
generated, so that this environment must be \emph{saved}, it is the role of the
closure technology.

We have now in our toolbox the necessary ingredients allowing us to implement
the product bifunctor.

 \bmp
 \bmpi\verb|> (setq monoid-product|\empi
 \bmpi\verb|    (lambda (monoid1 monoid2)|\empi
 \bmpi\verb|      (list (funcall monoid-product-identity monoid1 monoid2)|\empi
 \bmpi\verb|            (funcall monoid-product-operation monoid1 monoid2))))|\empim
 \bmpi\verb|#<Interpreted Function (unnamed) @ #x20f00c52>|\empix
 \bmpi\verb|> (setq N+_x_N* (funcall monoid-product N+ N*))|\empim
 \bmpi\verb|((0 1) #<Interpreted Closure (unnamed) @ #x20f06ae2>)|\empix
 \bmpi\verb|> (funcall operation N+_x_N*|\empim
 \bmpi\verb|           (list 4 4) (list 5 5))|\empi
 \bmpi\verb|(9 20)|\empi
 \emp

\noindent In other words, in the monoid \(\boxtt{N+} \times \boxtt{N*}\), the
product of \((4,4)\) and \((5,5)\) is \((9,20)\). Now these constructors can be
used for arbitrary monoids. For example the monoid \boxtt{CS} of the character
strings and their concatenation is constructed in Lisp as follows. Using the
generic function \boxtt{concatenate} of Lisp, a specific function
\boxtt{string-conc} is constructed, giving the second component of our monoid;
the identity is the empty string.

 \bmp
 \bmpi\verb|> (setq string-conc|\empi
 \bmpi\verb|  (lambda (string1 string2)|\empi
 \bmpi\verb|    (concatenate 'string string1 string2)))|\empim
 \bmpi\verb|#<Interpreted Function (unnamed) @ #x20ded9a2>|\empix
 \bmpi\verb|> (funcall string-conc "qwert" "yuiop")|\empim
 \bmpi\verb|"qwertyuiop"|\empix
 \bmpi\verb|> (setq CS (list "" string-conc))|\empim
 \bmpi\verb|("" #<Interpreted Function (unnamed) @ #x20ded9a2>)|\empix
 \bmpi\verb|> (funcall operation CS "qwert" "yuiop")|\empim
 \bmpi\verb|"qwertyuiop"|\empi
 \emp

The product \(\boxtt{CS} \times \boxtt{N+}\) is then constructed and tested:

 \bmp
 \bmpi\verb|> (setq CS_x_N+ (funcall monoid-product CS N+))|\empim
 \bmpi\verb|(("" 0) #<Interpreted Closure (unnamed) @ #x20e8ea6a>)|\empix
 \bmpi\verb|> (funcall operation CS_x_N+|\empi
 \bmpi\verb|         (list "qwert" 4) (list "yuiop" 5))|\empim
 \bmpi\verb|("qwertyuiop" 9)|\empix
 \emp

The product \(\boxtt{CS} \times (\boxtt{N+} \times \boxtt{N*})\) is constructed
in the same way.

 \bmp
 \bmpi\verb|> (setq CS_x_[N+_x_N*]|\empi
 \bmpi\verb|    (funcall monoid-product CS N+_x_N*))|\empim
 \bmpi\verb|(("" (0 1)) #<Interpreted Closure (unnamed) @ #x20eb02b2>)|\empix
 \bmpi\verb|> (funcall operation CS_x_[N+_x_N*]|\empi
 \bmpi\verb|                     (list "qwert" (list 4 5))|\empi
 \bmpi\verb|                     (list "yuiop" (list 4 5)))|\empim
 \bmpi\verb|("qwertyuiop" (8 25))|\empi
 \emp

The product bifunctor is present in our environment and can be used as in
category theory. Its main component is a functional object constructing a new
functional object from two other functional objects, the closure notion
allowing the user to naturally construct this higher functional object, for the
constructed closure ``keeps'' the environment where it has been generated. In
this way, arbitrary complex uses of the product bifunctor can be done whithout
any confusion: every closure will see the right references. \emph{And} the
complexity cost in this case is constant, only the addresses of \boxtt{monoid1}
and \boxtt{monoid2} are used; the code constituting the heart of the function
\boxtt{monoid-product-operation} being once for all installed in the
environment, which code can be used by \emph{any} product of monoids, thanks to
the indirect addressing mechanism.

\section{When computing homotopy groups.}

Computing the homotopy groups of a finite simply connected simplicial set, or
more generally a simply connected simplicial set \emph{with effective
homology}, using its Whitehead or Postnikov tower, is just the application of a
long sequence of applications of functors similar to the one used as example in
the previous section. For example the construction of \(\del_{X_i}\) from
\(X_{i-1}\) discussed in Section~\ref{wi} is obtained by such a long process,
and the same for the other components of \(X_i\)

 When the functional components of the initial objects are
polynomial, all the used functors are so simple that of course the resulting
objects obviously again have polynomial functional components.

But this is not enough: the ``measure of polynomiality'' of the output
functional objects must in turn polynomially depend on the measure of
polynomiality of the inputs.

The final scaffolding of functors is certainly a little impressive, but once
you master the notion of closure, the proof of polynomiality is a simple
recursive process.

We just give here a minimal significant example of the method to be applied.
Let \(A, B, C, D\) be four ``ordinary'' types of objects for which a size is
defined, such as integers, lists of such objects, matrices of such objects,
etc. Let \([A \rightarrow B]\) (resp. \([C \rightarrow D]\)) be the functional
types of the functional objects which, when an input is an element of \(A\)
(resp. \(C\)), return an element of \(B\) (resp. \(D\)). Let finally \(\alpha
\in [[A \rightarrow B] \rightarrow [C \rightarrow D]]\) be a sort of
``functor'' between these functional types. Question, how to define the
polynomiality of \(\alpha\)?

We would like to express that \emph{any} polynomiality of \(f: A \rightarrow
B\) implies \emph{some} polynomiality of \(\alpha(f): C \rightarrow D\), but in
a ``polynomial'' way. After a little work, you quickly find the appropriate
definitions are the following ones.

\begin{dfn} --- A functional object \(f: A \rightarrow B\) is an element of
\(P^d_c(A,B)\) if for any element \(a \in A\), the estimate \(\tau(f, a) \leq
c(1+\sigma(a))^d\) holds, where \(\sigma(a)\) is the size of the input \(a\)
and \(\tau(f, a)\) is the computing time of the output \(f(a)\).
\end{dfn}%

\begin{dfn} --- The functional object \(\alpha : [A
\rightarrow B] \rightarrow [C \rightarrow D]\) is polynomial if for every
degree \(d \in \bN\), there exists a degree \(d' \in \bN\) and a polynomial
\(\chi_d \in \bN[c]\) satisfying the following requirement: if \(f \in
P^d_c(A,B)\), then \(\alpha(f) \in P^{d'}_{\chi_\alpha(c)}\).
\end{dfn}%

In particular, no type of relation between \(d'\) and \(d\) is required, but a
\emph{unique} \(d'\) must be associated to a fixed \(d\); on the contrary the
``measure of \(d'\)-polynomiality'' of \(\alpha(f)\) must polynomially depend
on the ``measure of \(d\)-polynomiality'' of \(f\).

It is then obvious the composition of polynomial maps \(f_i: [A_{i-1}
\rightarrow B_{i-1}] \rightarrow [A_i \rightarrow B_i]\) is again polynomial.

The reader must be lucid: the definitions above concern the relations between
the respective complexities of \(\alpha: A \rightarrow B\) and \(f(\alpha): C
\rightarrow D\), and not the complexity of \(f\) itself. In the case of the
Kenzo program, this complexity is uniformly bounded by a constant and can
finally be omitted. This is the point never studied in~\cite{ckmvw}. In more
complex situations, the complexity of \(f\) could have a non-negligible role.

Also, you must handle more complicated cases where inputs and outputs are
mixtures of ``ordinary'' objects with an ``ordinary'' size and functional
objects of arbitrary level. There is an essentially unique way to extend the
simple case explained above to the general situation; it is not useful to
detail this point here.

In this way, the Kenzo algorithm can be mathematically described exactly as it
was programmed many years ago, directly following Postnikov; and the desired
polynomiality proof is then quite simple; and correct.

\section{Functional vs Ordinary programming.}

To be complete, it is probably useful to detail the exact status of
\emph{functional} programming with respect to ``ordinary'' programming.

The standard definitions of theoretical programming carefully distinguish the
\emph{primitive} recursive programs from the \emph{general} recursive programs;
necessarily illustrated by the example of the Ackermann function, which can be
programmed as a general recursive program, not as a primitive recursive
program. A reader of this text could wonder whether there exists such a
theoretical difference between functional programming and ordinary programming.

In a sense, this text proves there is no difference. The closure technology is
nothing but a process translating a functional program into an ordinary
program, so that you cannot program \emph{more} algorithms with functional
programming. But if you do not want to use functional programming, you will
have to manage by yourself the serious related problem of identifier scope. The
closure technology has precisely been invented to free the user of these low
level programming problems, without any cost in complexity, thanks to a skilful
use of indirect addressing\footnote{An amusing illustration of which happens
without closures is the funny ``solution'' used by Bourbaki~\cite{bourbk} to
solve this problem by the \emph{graphic} connections between Hilbert's
\(\tau\)'s and the white boxes \(\Box\).}.

The reader could usefully consult again the Wikipedia page ``Closure (computer
programming)''; it is there explained the so-called functional languages all
have integrated the notion of closure, in a so essential way that the keyword
\emph{closure} is even not present in these languages: their users employ
closures as Monsieur Jourdain wrote in prose.

In the same Wikipedia page, it is explained how in the other languages it is
possible to explicitly implement closures if necessary; the \emph{object
oriented programming} tool is then often used to solve the problem of
identifier scope, while in fact this tool is not at all devoted to this matter.
Once these techniques are understood, you should be able to translate the Lisp
Kenzo program into any other language, with a source code, for the
non-functional languages, certainly much less convenient; ask for example
\texttt{\small Google(compose1 c++)} to see the problems met by the
\boxtt{C++}-writers of the \emph{toy} example of Section~\ref{compose} and
imagine them translating the 16000 lines of Kenzo in \boxtt{C++}.

But using these (pseudo-)alternatives does not change anything to the problem:
anyway, functional objects have to be dynamically generated during the
execution of the program and you must control the cost of these generations. If
you can prove the cost of the generation of \emph{one} functional object is
uniformly bounded ---~be careful: independently of the \emph{size of the
initial object}~--- you have finished. Otherwise, you have to study this
dependency, good luck! You understand now how the closure technology is
powerful: in the case of the Kenzo program, it is easy to prove the cost of the
generation of a closure is uniformly bounded, independently of the initial
object.

Other techniques can be used. For example it can be proved the functional style
used in Kenzo can be replaced by a strong recursive style; in fact closer
(\(!\)) to the \emph{strict} theoretical functional style. More precisely, the
bottom-up programming style of Kenzo may be replaced by a top-down programming
style, producing a program where it \emph{seems} no functional object is
dynamically constructed. It is not at all convenient, would you like to start
the building of a house by the roof and finish by the foundations?  And this
would not change anything in the problem: this time you have to study the cost
of the handling of the recursiveness, not easy in this situation with a
terrible graph of crossed multi-recursiveness. And the ``upside-down''
understanding of the algorithm probably makes the complexity study rather
painful. Anyway it would be a \emph{different} algorithm.

In fact, the paper~\cite{ckmvw} does not give the smallest indication about the
solution of the authors for this problem. Following the Kenzo program, the
authors vaguely explain that for any simplicial set with effective homology
\(X\), an algorithm \(\pi^X_n\) can be written down computing \(\pi_n(X)\)
which is claimed uniformly polynomial with respect to \(X\), but the study of
the algorithm \(X \mapsto \pi^X_n\) is missing. No indication at all about
which is written before and after~(\(!\)) launching the execution, for example
about the Eilenberg-MacLane spaces to be \emph{generated}, with all their
\emph{functional} components, during the execution of \(w_i\), see
Section~\ref{wi}.

Also, trivially using functional programming, the Kenzo program is independent
of the \(n\) of \(\pi_n(X)\); nothing is said about this point in~\cite{ckmvw}.
It would be funny, for a process so simply iterative, to see a program
depending on the number of steps.

In summary, the cost of the generation of the countless functional objects is
not considered in the paper, while it is the heart of the program. Nothing is
explained either about a possible method to avoid functional programming.
Observe in particular the footnote~11 of~\cite{ckmvw}, which explains
``functional programming'' could be an ``alternative'' framework to process our
complexity problem ; the confusion is total: the program as it is described
in~\cite{ckmvw} must inevitably use functional programming; and ``functional
programming'' cannot be a \emph{tool} to study the complexity of an algorithm.

The article~\cite{ckmvw} reasonably updated could be a useful introduction to
the subject, but the claimed proof of polynomiality is incomplete. It is true
it is difficult to master this subject if you do not have any serious concrete
experience of programming. The same authors, along the same lines, have proved
an impressive collection of computability and undecidability results, results
which, as far as I know, are correct; opening, in the positive case, a
wonderful field for concrete programming. It is a pity to spoil the excellent
appreciation due to all these results by the paper~\cite{ckmvw}.

\bibliography{Francis}
\bibliographystyle{abbrv}

\end{document}